\documentclass[11pt]{article}
\usepackage{verbatim}
\usepackage{fullpage}
\usepackage{amsfonts}
\usepackage{amssymb}
\usepackage{color}
\usepackage{graphicx}
\usepackage{helvet}
\usepackage{latexsym}
\usepackage{amsgen}
\usepackage{amsmath}
\usepackage{amsopn}
\usepackage{amsxtra}

\newtheorem{theorem}{Theorem}[section]
\newtheorem{lemma}[theorem]{Lemma}
\newtheorem{proposition}[theorem]{Proposition}

\newtheorem{conjecture}[theorem]{Conjecture}
\newtheorem{algorithm}[theorem]{Algorithm}

\newenvironment{proof}[1][Proof:]{\begin{trivlist}
\item[\hskip \labelsep {\bfseries #1}]}{\end{trivlist}}

\newcommand{\qed}{\hfill\rule{2mm}{2mm}}

\newcommand{\ket}[1]{|#1\rangle}
\newcommand{\braket}[2]{\langle #1|#2\rangle}

\newcommand{\B}{\mathcal{B}}
\newcommand{\D}{\mathcal{D}}

\renewcommand{\H}{\mathcal{H}}

\newcommand{\R}{\mathbb{R}}

\newcommand{\Z}{\mathbb{Z}}
\newcommand{\W}{\mathcal{W}}

\newcommand{\C}{\mathbb{C}}

\newcommand{\mix}{\operatorname{mix}}
\newcommand{\hit}{\operatorname{hit}}
\newcommand{\diam}{\operatorname{diam}}

\begin{document}

\title{Almost uniform sampling via quantum walks\thanks{This material
is based upon work supported by the National Science Foundation under
Grant No. 0523866.}}
\author{Peter C. Richter\thanks{Department of Computer Science,
Rutgers University, Piscataway, NJ 08854.  {\tt richterp@cs.rutgers.edu}}}
\date{}
\maketitle

\begin{abstract}
Many classical randomized algorithms (e.g., approximation algorithms
for \#P-complete problems) utilize the following random walk algorithm for
{\em almost uniform sampling} from a state space $S$ of cardinality
$N$: run a symmetric ergodic Markov chain $P$ on $S$ for long enough
to obtain a random state from within $\epsilon$ total variation distance of the
uniform distribution over $S$.  The running time of this algorithm, the
so-called {\em mixing time} of $P$, is $O(\delta^{-1} (\log N + \log
\epsilon^{-1}))$, where $\delta$ is the spectral gap of $P$.

We present a natural quantum version of this algorithm based on
repeated measurements of the {\em quantum walk} $U_t = e^{-iPt}$.  We
show that it samples almost uniformly from $S$ with logarithmic
dependence on $\epsilon^{-1}$ just as the classical walk $P$ does; previously, 
no such quantum walk algorithm was known.  We then outline a framework for
analyzing its running time and formulate two plausible conjectures
which together would imply that it runs in time $O(\delta^{-1/2}
\log N \log \epsilon^{-1})$ when $P$ is the standard transition matrix
of a constant-degree graph.  We prove each conjecture for a subclass
of Cayley graphs.
\end{abstract}


\section{Introduction}

\subsection{Quantum walks and algorithms}

In the design of quantum algorithms, we are interested in identifying
problems whose time complexity appears to drop significantly if we
allow solution by a quantum computer.  Notable examples discovered
thus far include the hidden subgroup problem (Simon \cite{Sim},
Shor \cite{Sho}, Kitaev \cite{Kit}, Hallgren \cite{Hal}) and the
unstructured search problem (Grover \cite{Gro}).  While the apparent
quantum speedup is exponential for the hidden subgroup algorithms
(which use the powerful quantum Fourier transform), the quadratic
speedup of Grover's search algorithm is significant in that it
is transferrable to a broad array of problems, as the search problem
is perhaps the most fundamental in computer science.

Over the last several years, a family of algorithms based on
{\em quantum walks} (see the surveys by Kempe \cite{Kem} and Ambainis
\cite{Amb2}) has emerged as a considerable generalization of
Grover's algorithm.  A quantum walk is the analogue of a classical
random walk in that it is the powering of a single, often sparse,
matrix; in the quantum case this matrix is unitary rather than
stochastic.  Discrete-time quantum walks were first investigated in
the computer science community by Meyer \cite{Mey} and
Watrous \cite{Wat}, then more explicitly by Nayak et al.
\cite{NV,ABNVW} and Aharonov et al. \cite{AAKV}.  They have been used
in quantum algorithms for the element distinctness problem
(Ambainis \cite{Amb1}), matrix product verification (Burhman and
Spalek \cite{BS}), finding triangles in graphs (Magniez, Santha, and
Szegedy \cite{MSS}), finding subsets (Childs and Eisenberg \cite{CE}),
and group commutativity testing (Magniez and
Nayak \cite{MN}).  Continuous-time quantum walks were introduced by
Childs, Farhi, and Gutmann \cite{FG,CFG}.  They were used by Childs et
al. \cite{CCD} to solve in polynomial time an oracle problem for which
no polynomial-time classical algorithm exists.\footnote{Continuous-time
quantum walks can be simulated efficiently by (discrete-time) quantum
circuits using techniques which we discuss in Section \ref{simulability}.}

\subsection{Quantum hitting and mixing times}

Each of the quantum walk algorithms we have just mentioned solves a
particular instance of the following abstract {\em search} problem:
given a graph structure (i.e., a state space with allowed transitions)
and a subset of states which are {\em marked}, find a marked state (if
one exists) by performing a quantum walk on the graph.
In the {\em decision} version of the problem, we need only detect
with constant success probability whether or not the marked subset
is nonempty; we call its complexity (measured in walk steps) the
{\em quantum hitting time}.\footnote{The search and decision versions
for classical random
walks have the same complexity; this does not appear to be true for
quantum walks.}  The discrete- and continuous-time versions of the
quantum hitting time on various graphs have been studied by a number of
researchers (see e.g., \cite{Amb1,Kem2,SKW,AKR,CG1,CG2,Chi}).  Szegedy
\cite{Sze1,Sze2} extended the notion of discrete-time quantum
hitting time to ``quantized'' Markov chains and showed the following
general result: given a symmetric ergodic Markov chain with classical
hitting time $\tau_{\hit}$, we can define a natural unitary quantum
walk with the same locality whose quantum hitting time is
$O(\tau_{\hit}^{1/2})$.  In particular, the classical hitting time of
a Markov chain with spectral gap $\delta$ and a fraction $\varepsilon$
of its states marked is $O((\delta \varepsilon)^{-1})$, so the quantum
hitting time is $O((\delta \varepsilon)^{-1/2})$.

For classical Markov chains, the {\em mixing time} parameter is as
significant in applications as the hitting time, if not more so.  The
mixing time of an ergodic Markov chain measures the time past which
the current state distribution is within $\epsilon$ distance of its (limiting)
stationary distribution over all states, independent of what the initial
state was.  Thus, the mixing time characterizes the complexity of
{\em sampling} from the states of a Markov chain with respect to the
stationary distribution.  The most famous application in theoretical computer
science is in the Markov chain Monte Carlo (MCMC) method for
approximate sampling and counting.  An important special case is the
problem of {\em almost uniform sampling} (see Jerrum \cite{Jer}),
where we wish to output an element from within $\epsilon$ total variation
distance of the uniform distribution over a finite set, in this case
by running a symmetric Markov chain whose stationary distribution is uniform
over the set until it has $\epsilon$-mixed.

\subsection{Our contributions}

In light of the success of quantum walks in speeding up classical
search algorithms, an obvious question to ask is: can classical sampling
algorithms based on mixing of random walks be sped up by using quantum
walks instead?  Aharanov et al. \cite{AAKV} were the first to look
closely at the subject of (time-averaged) limiting distribution for a
quantum walk and to
define a notion of {\em quantum mixing time} to reach this
distribution.  A number of important questions were raised in their
paper.  We consider two in particular, the first being: which
distributions can we generate using quantum walks?  Aharanov et
al. \cite{AAKV} showed that the limiting
distribution of the quantum walk on a regular graph is typically
non-uniform (and even dependent on the initial state); we give a
simple algorithm based on repeated measurements of the continuous-time
quantum walk $U_t = e^{-iPt}$ for a symmetric ergodic Markov chain $P$
(inspired by the decohering quantum walk of Kendon and Tregenna
\cite{KT,Ken}) that necessarily outputs a state $\epsilon$-close to
the uniform distribution.  Moreover, it does so with only logarithmic
dependence on $\epsilon^{-1}$, just as the classical walk $P$ does.

The second question of Aharanov et al. \cite{AAKV} that we consider
is: when does quantum mixing offer a speedup over
classical mixing?  This question seems to be a great deal more
difficult than the first (e.g., Aharonov et al. \cite{AAKV} only
demonstrated a quantum speedup for the discrete-time walk on a cycle);
we provide a framework for answering this question and then formulate
two strong but plausible conjectures: (a) that {\em threshold quantum
mixing}  can be achieved in $O(\delta^{-1/2} \log N)$ time when $P$ is
the standard transition matrix of a regular graph, and (b) that
{\em amplification} to $\epsilon$-mixing from threshold-mixing can be
achieved in $O(\log \epsilon^{-1})$ time ({\em independent} of $N$)
when $P$ is the standard transition matrix of a constant-degree
graph.  In particular, if both conjectures are true, then there is an
$O(\delta^{-1/2} \log N \log \epsilon^{-1})$ quantum algorithm for
almost uniform sampling from the vertices of a constant-degree graph,
as opposed to the $O(\delta^{-1} (\log N + \log \epsilon^{-1}))$
classical algorithm.  We prove the amplification conjecture for the
torus $\Z_p^d$, $p$ prime (the proof is more subtle than one might
expect), and we prove the threshold mixing conjecture for the
hypercube $\Z_2^n$ and the complete graph $K_N$.  It is
our hope that our approach and progress on the problem of almost
uniform sampling via quantum walks will rejuvenate interest in
characterizing the speedup of quantum mixing.

\section{Preliminaries}\label{preliminaries}

\subsection{Classical and quantum processes}

Let $S$ be a set of {\em states} with $|S|=N$.  Consider the
$N$-dimensional complex vector space $\H(S)$ with {\em basis states}
$\B(S) := \{\ket{s} : s \in S\}$; in particular, this space contains
the {\em pure quantum states} (i.e., wavefunctions) $\W(S) :=
\{\ket{\phi} = \sum_{s \in S} \phi_s \ket{s} : \phi_s \in \C \:
\forall s \in S, ||\phi||_2 = 1\}$.  It also contains the
{\em classical randomized states} (i.e., distributions) $\D(S) :=
\{p = \sum_{s \in S} p_s \ket{s} : p_s \in \R_{\geq 0} \: \forall s
\in S, ||p||_1 = 1\}$.

For our purposes, it will suffice to consider the three following
physically realizable processes: {\em stochastic evolution} $\D(S)
\rightarrow \D(S)$ (e.g., a symmetric Markov chain $P$), {\em unitary
evolution} $\H(S) \rightarrow \H(S)$ (e.g., the {\em quantum walk} $U_t
: \ket{\phi} \mapsto e^{-iPt} \ket{\phi}$), and the {\em projective
measurement} $\H(S) \rightarrow \D(S) : \ket{\phi} \mapsto \sum_{s \in
S} \ket{s} \braket{s}{\phi}$.\footnote{Note that the ``missing'' process
$\D(S) \rightarrow \H(S)$ is unnecessary, since it can be described
as the linear combination of its (non-interfering) actions on basis
states $\B(S)$, which are a special case of unitary evolution.}

{\em Remark.}  Processes with quantum operations and intermediate
measurements produce {\em mixed states} (i.e., randomized ensembles
of quantum states) and are typically described by their action on
{\em density matrices}; however, our algorithm and its analysis are
more simply described using classical and quantum state vectors alone.

\subsection{Basic properties of Markov chains}

The long-term behavior of a discrete-time Markov chain $P$ and its
continuous-time counterpart (Poisson average) $\exp(-(I-P)t) =
\sum_{s=0}^\infty \frac{e^{-t} t^s}{s!} P^s$ is summarized by the
following theorems.\footnote{Throughout this paper, we use the term
Markov chain when describing either the actual chain ($\{P^t\}$ in
discrete time, $\{\exp(-(I-P)t)\}$ in continuous time) or the
stochastic matrix $P$ generating the chain.}

A {\em stationary distribution} of a Markov chain $P$ is a
distribution $\pi$ satisfying $P \pi = \pi$.
\begin{theorem}[Unique stationary distribution]
Let $P$ be {\em irreducible} (i.e., strongly connected).  Then there
exists a unique stationary distribution $\pi$ for $P$.
\end{theorem}
In particular, if $P$ is symmetric then the uniform distribution $u :=
[\frac{1}{N}, \ldots, \frac{1}{N}]^\dagger = \frac{1}{N} 1^\dagger$ is
stationary.\footnote{We use $1^\dagger$ to denote the $N$-dimensional
row vector of ones; $\dagger$ is the conjugate transpose.}

An irreducible Markov chain $P$ converges in the {\em Cesaro} (i.e.,
time-averaged) and continuous-time (i.e., Poisson-averaged) limits to
its stationary distribution:
\begin{theorem}[Fundamental theorem]
Let $P$ be irreducible.  Then the stochastic matrices
$\frac{1}{T}\sum_{t=0}^{T-1} P^t$ and
$\sum_{s=0}^\infty \frac{e^{-T} T^s}{s!} P^s$ tend to $[\pi \pi \cdots
\pi] = \pi 1^\dagger$ as $T \rightarrow \infty$.  If $P$ is also {\em
aperiodic} (i.e., non-bipartite), then $P^t \rightarrow \pi 1^\dagger$
as $t \rightarrow \infty$, and we call $P$ {\em ergodic}.
\end{theorem}

An irreducible Markov chain is {\em reversible} if it satisfies the
{\em detailed balance} constraints $P_{ij} \pi_j = P_{ji} \pi_i$; or
equivalently, if the matrix $S = D P D^{-1}$ is symmetric, where $D$
is the diagonal matrix with $D_{ii} = \sqrt{\pi_i}$.  The {\em
spectral gap} $\delta$ of a Markov chain $P$ is the difference between
$1$, its largest eigenvalue, and $\lambda$, its second-largest
eigenvalue (in absolute value).  The {\em mixing time}
$\tau(\epsilon)$ of an ergodic discrete-time Markov chain is $\min \{
T : \frac{1}{2} ||P^t - \pi 1^\dagger||_1 \leq \epsilon \: \forall t
\geq T \}$, where $\frac{1}{2}||\cdot||_1$ is the {\em total
variation} distance.
\begin{theorem}[Diaconis, Strook \cite{DS}; Aldous \cite{Ald}]\label{ds-ineq}
Let $P$ be reversible and ergodic.  Then its mixing time satisfies
$\frac{1}{2}\lambda \delta^{-1} \ln(2 \epsilon)^{-1} \leq
\tau(\epsilon) \leq \delta^{-1} (\ln \pi_{\min}^{-1} + \ln \epsilon^{-1})$,
where $\pi_{\min} := \min_x \pi(x)$.\footnote{We have implicitly
assumed here that the second-largest eigenvalue of $P$ exceeds the
smallest in absolute value; this is easy enough to arrange in practice
by replacing $P$ with $\frac{1}{2}(I+P)$.}
\end{theorem}
The mixing time of any irreducible Markov chain $P$ can be defined
using the Cesaro average, the Poisson average, or the lazy chain
$\frac{1}{2}(I+P)$; in each case, the mixing time behaves similarly up to
the dependence on $\epsilon$.  In most applications the key factor in
Theorem \ref{ds-ineq} is the spectral gap.  Sometimes we can compute
this directly; other times we estimate it, say via conductance (see
e.g., \cite{JSV,Jer}).

Theorem \ref{ds-ineq} implies that the following single-loop
classical mixing algorithm samples from a distribution
$\epsilon$-close to $\pi$ for any ergodic $P$, from any initial state:
\begin{algorithm}[Single-loop classical]
Do $\tau(\epsilon)$ times:  Apply $P$ to the current state.
\end{algorithm}

In describing our quantum algorithm for almost uniform sampling, we
will appeal to the notions of {\em theshold mixing} and
{\em amplification}, which are implicit in the classical Markov chain
mixing algorithm as well.  In particular, a (slightly weaker) variant
of the previous theorem follows from combining the next two
well-known theorems:

The {\em threshold mixing time} $\tau_{\mix}$ of an irreducible Markov
chain is $\tau_{\mix} := \tau(1/2e)$.
\begin{theorem}[Amplification]
For any Markov chain $P$, $\tau(\epsilon) \leq \tau_{\mix} \lceil \ln
\epsilon^{-1} \rceil$.
\end{theorem}
\begin{theorem}[Threshold mixing - Aldous \cite{Ald}]
Let $P$ be reversible and ergodic.  Then
$\delta^{-1} \leq \tau_{\mix} \leq \delta^{-1} (1 + \frac{1}{2} \log
\pi_{\min}^{-1})$.
\end{theorem}

In light of the amplification and threshold mixing theorems, we can
view the same classical mixing algorithm as a double-loop
algorithm instead: the inner loop
builds the fast-mixing threshold matrix $P^{\tau_{\mix}}$ from $P$,
and the outer loop iterates this matrix only $\lceil \ln
\epsilon^{-1} \rceil$ times to achieve $\epsilon$-mixing.
\begin{algorithm}[Double-loop classical]
Do $\lceil \ln \epsilon^{-1} \rceil$ times: \{ Do $\tau_{\mix}$ times:
Apply $P$ to the current state. \}
\end{algorithm}

\subsection{Additional properties of Markov chains}

We will use several more facts about Markov chains and their mixing
properties beyond the basic ones already mentioned; we present them
here with proofs for completeness.

Why do we set the threshold mixing time $\tau_{\mix}$ to $\tau(1/2e)$?
We do it precisely so that the amplification theorem holds (i.e., so
that $P^{\tau_{\mix}}$ $\epsilon$-mixes in time $O(\log \epsilon^{-1})$).
It so happens that the choice $1/2e$ is rather arbitrary beyond the fact
that it is a fixed constant below $1/2$, as the following
generalization of the amplification theorem shows.
\begin{proposition}\label{pairwise}
Let $Q$ be a Markov chain with maximum pairwise column distance $\max_{x,x'}
\frac{1}{2}||Q(\cdot,x)-Q(\cdot,x')||_1 \leq \alpha < 1$.  Then it has
mixing time
$\tau(\epsilon) \leq \lceil \log_{1/\alpha} \epsilon^{-1} \rceil$.
\end{proposition}
\begin{proof}
From Aldous and Fill \cite{AF}: Let $\bar{d}(t) := \max_{x,x'}
\frac{1}{2}||Q^t(\cdot,x)-Q^t(\cdot,x')||_1$ be the maximum pairwise
column distance at time $t$.  This distance is submultiplicative;
i.e., $\bar{d}(s+t) \leq \bar{d}(s)\bar{d}(t)$ for any $s,t \geq 0$.
In particular, $\bar{d}(t) \leq \bar{d}(1)^t$, so $d(\lceil
\log_{1/\alpha} \epsilon^{-1} \rceil) \leq \bar{d}(\lceil
\log_{1/\alpha} \epsilon^{-1} \rceil) \leq \epsilon$, where $d(t) :=
\frac{1}{2}||Q^t - \pi 1^\dagger||_1$ and $\pi$ is the stationary
distribution of $Q$.\qed
\end{proof}
In particular, if $T \geq \tau(1/2e)$, then the columns of $P^T$ are
$(1/2e)$-close to their common limit $\pi$, so they have maximum
pairwise distance at most $1/e$.  It follows that $P^T$ mixes in time
$\lceil \log_e \epsilon^{-1} \rceil = \lceil \ln \epsilon^{-1}
\rceil$, yielding the amplification theorem.

So in order to show that a Markov chain mixes in time $O(\log
\epsilon^{-1})$, it suffices to show that its columns have maximum
pairwise distance at most some constant less than one.  A sufficient
condition for this property to hold is the following:
\begin{proposition}\label{entry-lb}
Let $\beta > \frac{1}{2}$, $\gamma > 0$; let $Q$ be an $N \times N$ stochastic
matrix with at least $\beta N$ entries in each column bounded below by
$\gamma / N$.  Then $\max_{x,x'}
\frac{1}{2}||Q(\cdot,x)-Q(\cdot,x')||_1 \leq 1-\gamma(1-2(1-\beta)) < 1$.
\end{proposition}
\begin{proof}
Recall that for any two distributions $p,q$ we have
$\frac{1}{2}||p-q||_1 = 1 - \sum_k \min\{p_k,q_k\}$.  It follows that
$\max_{x,x'}\frac{1}{2}||Q(\cdot,x)-Q(\cdot,x')||_1 \leq
1-(1-2(1-\beta))N \cdot \gamma/N = 1 - \gamma(1-2(1-\beta))$.\qed
\end{proof}

We can estimate the maximum pairwise column distance of a Markov chain
$Q'$ from that of a nearby chain $Q$ using the triangle inequality:
\begin{proposition}\label{perturbation}
Let $Q$ and $Q'$ be Markov chains; let $\beta := \frac{1}{2}||Q-Q'||_1$
and $\gamma := \max_{x,x'} \frac{1}{2} ||Q(\cdot,x)-Q(\cdot,x')||_1$.
Then $\max_{x,x'} \frac{1}{2} ||Q'(\cdot,x) - Q'(\cdot,x')||_1 \leq
2\beta + \gamma$.
\end{proposition}
\begin{proof}
For any $x,x'$: $\frac{1}{2}||Q'(\cdot,x) - Q'(\cdot,x')||_1 \leq
\frac{1}{2}||Q'(\cdot,x) - Q(\cdot,x)||_1 + \frac{1}{2} ||Q'(\cdot,x') -
Q(\cdot,x')||_1 + \frac{1}{2} ||Q(\cdot,x) - Q(\cdot,x')||_1$.
\qed
\end{proof}

\section{A quantum algorithm for almost uniform sampling}\label{algorithm}

\subsection{A single-loop algorithm}

Let $P$ be an irreducible reversible Markov chain with uniform
stationary distribution on the set $S$ of states;
equivalently, $P$ is an irreducible symmetric Markov chain.  We wish
to sample almost uniformly from $S$ using the quantum walk
$U_t = \exp(-iPt)$ (i.e., $P$ is the time-independent
{\em Hamiltonian}, or complex Hermitian matrix, driving the walk) from
any initial basis state $\ket{x} \in \B(S)$.
Aharonov et al. \cite{AAKV} propose (a discrete-time version of)
the following ``single-loop'' algorithm:

\begin{algorithm}[Single-loop quantum]
Run the walk $U=e^{-iPt}$ for time $t$ chosen uniformly at random
(u.a.r.) from $[0,T]$, then measure and output the current state.
\end{algorithm}

Let $P_t(y,x) := |\braket{y}{e^{-iPt}|x}|^2$ be the stochastic matrix mapping
inputs to outputs for this algorithm for a fixed $t$; then the
{\em finite-time Cesaro matrix} $\bar{P}_T := \frac{1}{T}\int_{t=0}^T
P_t \: dt$ maps inputs to outputs for $t$ chosen u.a.r. from $[0,T]$.
Unlike $P_t$ (whose columns oscillate as $t \rightarrow \infty$),
$\bar{P}_T$ has a limit $\Pi$ as $T \rightarrow \infty$, which we
call the {\em infinite-time Cesaro matrix}.  This is demonstrated by
the following continuous-time version of Theorem 3.4 from Aharonov et
al. \cite{AAKV}, which also gives an upper bound on the {\em quantum
mixing time} $\tau'(\epsilon) := \min\{T' : \frac{1}{2}||\bar{P}_T -
\Pi||_1 \leq \epsilon \}$:\footnote{Aharonov et al. \cite{AAKV} use a
slightly different notion of quantum mixing time than we do here.}

\begin{theorem}[Cesaro matrices]\label{spectral-ct}
Let $H$ be a Hamiltonian with spectrum $\{\lambda_k,\ket{\phi_k}\}$.
Let $\{C_j\}$ be the partition of these indices $k$ obtained by
grouping together the $k$ with identical $\lambda_k$.  Then:
\begin{eqnarray}\label{tcesaro-ct}
\bar{P}_T(y,x) & = & \sum_j \bigl| \sum_{k \in C_j}
\braket{y}{\phi_k}\braket{\phi_k}{x} \bigr|^2 \\
& + & \sum_{k,l : \lambda_k \neq \lambda_l} \braket{\phi_k}{x}
\braket{x}{\phi_l} \braket{y}{\phi_k} \braket{\phi_l}{y}
\frac{1}{T}\int_{t=0}^T e^{i(\lambda_l-\lambda_k)t} dt
\end{eqnarray}
As $T \rightarrow \infty$ the latter term tends to zero, so the limit
$\Pi$ of $\bar{P}_T$ exists and is equal to the former term.  Hence
the quantum mixing time $\tau'(\epsilon)$ is the smallest $T$ such
that:
\begin{equation}\label{mixing-ct}
\frac{1}{2}||\bar{P}_T - \Pi||_1 = \max_x \sum_y
\bigl|\sum_{k,l : \lambda_k \neq \lambda_l} \braket{\phi_k}{x}
\braket{x}{\phi_l} \braket{y}{\phi_k} \braket{\phi_l}{y}
\frac{1}{T}\int_{t=0}^T e^{i(\lambda_l-\lambda_k)t} dt \bigr|
\leq \epsilon
\end{equation}
\end{theorem}

It can be inferred from this theorem that both the discrete- and
continuous-time walks typically lack the property $\Pi = u 1^\dagger =
\frac{1}{N} \times \textrm{ the all-ones matrix }$
(which we require to mix to uniform, or even to mix to the same
distribution from any two initial states, using the above algorithm),
except in special circumstances such as when the walk takes place on
the Cayley graph of an Abelian group and it has distinct eigenvalues.
It so happens that these two special properties hold for the
discrete-time walk on the cycle (with Hadamard coin flip); Aharonov et
al. \cite{AAKV} show this and then use Theorem \ref{spectral-ct} to
prove that its quantum mixing time is $O(N \log N \epsilon^{-3})$,
yielding a quantum algorithm for almost uniform sampling on the cycle
that is nearly quadratically faster than the classical simple random
walk.  They further show that walks with the property
$\Pi = u 1^\dagger$ can be sequentially repeated in such a way that the
distance of the output state from uniform drops exponentially with the
number of walk repetitions (i.e., the dependence on $\epsilon^{-1}$ is
logarithmic), just like it does for classical Markov chains.  In particular,
this reduces the time required to sample almost uniformly from the
cycle to $O(N \log N \log \epsilon^{-1})$.

\subsection{A double-loop algorithm}

What about the (typical) case when $\Pi \neq u
1^\dagger$?\footnote{For a walk on the symmetric group $S_n$, Gerhardt
and Watrous \cite{GW} showed that $\frac{1}{2}||\Pi - u 1^\dagger||_1
\geq \frac{1}{n}-\frac{1}{n \cdot n!}\binom{2n-2}{n-1}$; for a walk on
the hypercube $\Z_2^n$, Moore and Russell \cite{MR} showed that there
exists an $\epsilon > 0$ such that $\frac{1}{2}||\Pi - u 1^\dagger||_1
\geq \epsilon$.}  Then not only is it unclear how to obtain the
exponential drop in closeness to uniform (i.e., logarithmic dependence
on $\epsilon^{-1}$), but the single-loop algorithm does not even sample almost
uniformly, but rather from a non-uniform distribution dependent on the
initial state.  To remedy this, we propose the following ``double-loop''
algorithm:

\begin{algorithm}[Double-loop quantum]
Do $T'$ times: Run the walk $U=e^{-iPt}$ for time $t$ chosen
u.a.r. from $[0,T]$, then measure and output the current state.
\end{algorithm}

This is almost the same algorithm proposed by Aharanov et
al. \cite{AAKV} to obtain logarithmic dependence on $\epsilon^{-1}$ in
the special case $\Pi = u 1^\dagger$, but we claim that for {\em any}
$\Pi$ our algorithm: (a) samples almost uniformly, and (b) does so with
logarithmic dependence on $\epsilon^{-1}$.

Let's prove claim (a):

\begin{theorem}[Uniform mixing]\label{pi-ct}
If $P$ is a symmetric irreducible Markov chain, then each entry
of $\Pi$ is bounded from below by $1/N^2$; in particular, $\Pi$ is
ergodic.  Moreover, each of the $P_t$ (and so $\bar{P}_T$ and $\Pi$
as well) is symmetric and therefore has
uniform stationary distribution.\footnote{$\Pi$ is also positive
semidefinite: it is the Gram matrix of $\{f_s\}$ with
$f_s(kl) := \braket{s}{\phi_k} \braket{\phi_l}{s}$ if $\lambda_k =
\lambda_l$, $0$ otherwise.}
\end{theorem}
\begin{proof}
The $1$-eigenspace of $P$ is precisely the space spanned by $u$, so it
follows from Theorem \ref{spectral-ct} that $\Pi(y,x) \geq 1/N^2$ for every
$x,y$.  To see that $P_t(y,x) := |\braket{y}{e^{-iPt}|x}|^2$ is
symmetric, write out the Taylor series for $e^{-iPt}$ and note that
every positive integer power $P^k$ is symmetric (since $P^2(x,y) =
\sum_z P(x,z) \cdot P(z,y) = \sum_z P(y,z) \cdot P(z,x) =
P^2(y,x)$).  It follows that the stationary distribution of $P_t$ is
uniform.\footnote{More generally, the uniform
distribution is stationary for any process consisting of unitary
evolution followed by measurement, since it is invariant under both
operations.}\qed
\end{proof}
This implies that for any $\epsilon > 0$ our algorithm will (for $T$
and $T'$ sufficiently large) output a state $\epsilon$-close to
uniform, so claim (a) is proven.

Now let's prove claim (b), and in the process obtain upper bounds on
the minimum $T$ and $T'$ required for our algorithm to output a state
$\epsilon$-close to uniform.  First a few definitions:  For a particular
quantum walk $e^{-iPt}$, fix $\alpha := \max_{x,x'} \frac{1}{2}
||\Pi(\cdot,x) - \Pi(\cdot,x')||_1$.  We define the {\em quantum
threshold mixing time} of the walk to be $\tau'_{\mix} :=
\tau'(\epsilon_0)$, where (for the time being) we set $\epsilon_0 :=
\frac{1-\alpha}{4}$.  Then we have the following easy theorem:

\begin{theorem}[Convergence time]\label{runtime-ct}
For $T = \tau'_{\mix}$ and $T' = \lceil \log_{2/(1+\alpha)}
\epsilon^{-1} \rceil$, our algorithm outputs a state $\epsilon$-close
to uniform.
\end{theorem}
\begin{proof}
If $T = \tau'_{\mix}$, then the Markov chain $\bar{P}_T$ built by the
inner loop is $\epsilon_0$-close to its limit $\Pi$, where $\epsilon_0 =
\frac{1-\alpha}{4}$.  By definition, the maximum pairwise column
distance of $\Pi$ is at most $\alpha$.  So by Proposition
\ref{perturbation}, $\bar{P}_T$ has maximum pairwise column distance
at most $2\epsilon_0 + \alpha = \frac{1+\alpha}{2}$.  Since the outer
loop is a $T'$-fold repetition of $\bar{P}_T$, choosing $T' \geq \lceil
\log_{2/(1+\alpha)} \epsilon^{-1} \rceil$ yields an output state
within $\epsilon$ distance of uniform by Proposition \ref{pairwise}.\qed
\end{proof}

We can now state clearly what we hope to be able to do: run the
quantum walk (inner loop) for a long enough time $T$ so that the
matrix $\bar{P}_T$ is within a small (but still constant) threshold
distance $\epsilon_0$ from its limit $\Pi$, at which point it must
have maximum pairwise column distance less than some constant below
one (provided that $\Pi$ itself has this property) and therefore need
only be applied $T' = O(\log \epsilon^{-1})$ times ({\em independent}
of $N$) to output a state $\epsilon$-close to uniform.  In particular,
this motivates us to make the next two strong but plausible
conjectures for the quantum walk $U_t = e^{-iPt}$:
\begin{conjecture}[Amplification]
Let $\alpha_0$ be the supremum of $\alpha$ over all quantum walks
for which $P$ is the standard transition matrix of a constant-degree
graph; then $\alpha_0$ is strictly less than one.
\end{conjecture}
This would imply that we can always choose $T'= O(\log \epsilon^{-1})$
for such walks.  Of course, this matters little unless the quantum
threshold mixing time is significantly faster than the classical
threshold mixing time; to this end, we conjecture:
\begin{conjecture}[Threshold mixing]
Any quantum walk for which $P$ is the standard transition matrix of a
regular graph satisfies $\tau'_{\mix} \leq O(\delta^{-1/2} \log N)$.
\end{conjecture}
This would imply that we can always choose $T = O(\delta^{-1/2} \log N)$
for such walks.  It would also imply (by Cheeger's inequality) that
the threshold quantum mixing time is $\tilde{O}(1/\Phi)$ where $\Phi$
is the conductance of $P$, a question posed by Aharonov et
al. \cite{AAKV}.  By Theorem \ref{runtime-ct}, the validity of both
conjectures would give us an $O(\delta^{-1/2} \log N \log
\epsilon^{-1})$ quantum algorithm for almost uniform sampling from the
vertices of a constant-degree graph.

\section{Proving the amplification and threshold mixing conjectures}
\label{conjectures}

\subsection{Intuition}\label{intuition}

To show that the amplification conjecture holds for a particular $P$,
we need an upper bound on the maximum pairwise column distance of
$\Pi$.  One way to do this is via Proposition \ref{entry-lb}, which
tells us that it suffices to show that most of the entries in each
column of $\Pi$ are $\Omega(1/N)$.  We can view this as a statement
that the quantum walk is ergodic in a weak sense; i.e., that from any
initial basis state, the limit of the time-averaged distribution
over states induced by the quantum walk is roughly equivalent to the
space-averaged (uniform) distribution.  Indeed, for Proposition
\ref{entry-lb} not to hold, almost all of the mass from the
time-averaged quantum walk distribution must be localized over a
minority of the states.  Using Theorem \ref{spectral-ct}, we have
already shown a lower bound of $1/N^2$ on each entry of $\Pi$ (Theorem
\ref{pi-ct}).  More detailed knowledge of the spectrum of $P$ allows
us to tighten this lower bound, as we shall soon see in the case of
the walk on the torus.

We remark that there are regular graphs of {\em non-constant} degree
for which the amplification conjecture fails to hold.  In particular,
the standard walk on the complete graph $K_N$ with self-loops mixes
classically in one step, but its quantum counterpart takes time
$\Theta(N \log \epsilon^{-1})$.  (This observation is due to
Fr\'{e}d\'{e}ric Magniez.)  Indeed, the squared amplitude of each
off-diagonal entry of $U_t = e^{-iPt}$ remains $O(1/N^2)$ for the
duration of the walk:
\begin{equation}
e^{-iPt} = \sum_{s=0}^\infty \frac{(-it)^s}{s!} P^s = I +
P(\sum_{s=1}^\infty \frac{(-it)^s}{s!}) = I + P(e^{-it}-1)
\end{equation}
Since $0 \leq |e^{-it}-1| \leq 2$ for all $t$, we have
$|\braket{y}{e^{-iPt}|x}|^2 \leq \frac{4}{N^2}$ for all $x \neq y$ and
all $t$, so $\alpha = \Theta(1-1/N)$ for this walk.

To show that the threshold mixing conjecture holds for a particular
$P$, we need to know how fast the quantum walk propagates.  It is
known that the distribution induced by a quantum walk spreads
quadratically faster than the corresponding random walk distribution
on the line and on higher-dimensional lattices, for example; on the
other hand, it cannot spread asymptotically faster than the (already
optimally-spreading) random walk distribution on a bounded-degree
expander.  Motivated by these and other observations, we conjecture
that the threshold mixing time of the quantum walk on a regular graph
$G$ is bounded above by $O(\diam(G))$, where $\diam(G)$ is the
diameter of $G$; in particular, it is known that $\diam(G) =
O(\delta^{-1/2} \log N)$.  The threshold mixing conjecture is true for
the (discrete-time) quantum walk on the cycle (Aharonov et
al. \cite{AAKV}) and for the quantum walks on the hypercube and
complete graph, which we will discuss shortly.

\subsection{The torus $\Z_p^d$}

We prove the amplification conjecture for the $d$-dimensional torus
$\Z_p^d$, $p$ prime.  The next two lemmas were proven with the help of
Mario Szegedy.

\begin{lemma}[Eigenvalue multiplicities of $\Z_p^d$]\label{multiplicities}
Let $P$ be the standard transition matrix on $\Z_p^d$ with $d \geq 1$
fixed, $p > 4d$ prime, and $N=p^d$.  Then each of the $C_j$ consists
of all $l \in \Z_p^d$ equivalent to a single $k \in \Z_p^d$ up to
permutation and signing of the coordinates.
\end{lemma}
\begin{proof}
Let $\omega = e^{2 \pi i / p}$.  Suppose $\lambda_k = \lambda_l$ and
$l$ is not equivalent to $k$.  Then the vanishing sum
\begin{equation}
\sum_{j=1}^d \omega^{k_j} + \omega^{-k_j} - \omega^{l_j} -
\omega^{-l_j} = 0
\end{equation}
is {\em not} simply $2d$ vanishing sums of length two $\omega^{k_j} -
\omega^{k_j}$, $\omega^{-k_j} - \omega^{-k_j}$ over $j : 1 \leq j \leq
d$.  Observe that if we cannot decompose the above sum into length-two
vanishing sums {\em in precisely this way}, then we cannot do so {\em
at all} (since $\omega^{k_j} + \alpha = 0 \Rightarrow \alpha =
-\omega^{k_j}$ is a $2p$th root of unity but {\em not} a $p$th root of
unity).  Hence there exists a minimal vanishing sum of length $m$
($3 \leq m \leq 4d$) of roots of unity $\zeta_i$ (wlog, wma $\zeta_1 =
1$) of common order $r=p$.  By Theorem 5 of Conway and Jones
\cite{CJ}, $\sum_{s \: prime : \: s | r} (s-2) \leq m-2$; so we have $p-2
\leq 4d-2$.\qed
\end{proof}

\begin{lemma}[Eigenvector cancellations for $\Z_n^d$]\label{cancellations}
For any $y \in \Z_n^d$, there are $\Omega(n)$ different $x \in \Z_n$
satisfying $x y_i \bmod n \in [-n/8d,n/8d]$ for all $i : 1 \leq i \leq d$.
\end{lemma}
\begin{proof}
Consider the map $f : \Z_n \rightarrow\Z_n^d$ given by $x \mapsto
(xy_1,\ldots,xy_d)$.  Thinking of $\Z_n^d$ as being divided into
subgrids of side length $n/m$ (with one of them, $H_0$, centered at
$0$), it is clear that one such subgrid (call it $H$) must contain at
least $n/m^d$ of the points in the image of $f$.  Let $x' y$ be any of
the image points in $H$.  Then there are at least $n/m^d$ different
$x$ for which $(x-x')y_i \in [-n/m,n/m]$ for all $i$.  Let $m=8d$;
then there are at least $n/(8d)^d$ such $x$.\qed
\end{proof}

\begin{theorem}[Amplification for $\Z_p^d$]
Let $P$ be the standard transition matrix on $\Z_p^d$ with $d \geq 1$
fixed, $p$ prime, and $N=p^d$.  Then each entry of $\Pi$ is bounded
below by $\Omega(1/N)$, so the amplification conjecture is satisfied.
\end{theorem}
\begin{proof}
Label the spectrum of $P$ using indices $k \in \Z_p^d$ as follows:
\begin{equation}
\lambda_k := \frac{1}{d} \sum_{j=1}^d \cos(2 \pi k_j / p) \qquad
\ket{\phi_k} := \frac{1}{\sqrt{N}} \sum_{x \in \Z_p^d} e^{2 \pi i k
\cdot x / p} \ket{x}
\end{equation}
Since $\Z_p^d$ is vertex-transitive, it suffices to show that
$\Pi(y,0) = \Omega(1/N)$; or, from Theorem \ref{spectral-ct}:
\begin{equation}
\sum_j |\sum_{k \in C_j} e^{2 \pi i k \cdot y / p}|^2 = \Omega(N)
\end{equation}
This is a consequence of the preceding two lemmas: by Lemma
\ref{cancellations}, at least $(p/(8d)^d)^d = \Omega(N)$ of the $k \in
\Z_p^d$ are such that: (i) $k_j y_i \bmod p \in [-p/8d,p/8d] \: \forall
i,j$, thus (ii) $l \cdot y := \sum_i l_i y_i \bmod p \in [-p/8,p/8]$
for all $l$ equivalent to $k$ up to permutation and signing of the
coordinates.  So by Lemma \ref{multiplicities}, which in particular
shows that $|C_j| \leq 2^d d!$ for all $j$, it follows that
$(p/(8d)^d)^d/2^d d! = \Omega(N)$ of the $C_j$ give a sum of at least
$1$.\qed
\end{proof}

\subsection{The hypercube $\Z_2^n$ and complete graph $K_N$}

We now give a simple argument showing that the threshold mixing
conjecture holds for the hypercube $\Z_2^n$ and complete graph $K_N$.

\begin{lemma}[Periodicity]\label{periodicity}
Let $P$ be the standard transition matrix of a regular graph $G$.  If
$U_t = e^{-iPt}$ is periodic with period $O(\diam(G))$, then $P$
satisfies the threshold mixing conjecture.
\end{lemma}
\begin{proof}
Let $T$ be the period; then $\bar{P}_T = \Pi$.  So the threshold
mixing conjecture is satisfied: $\tau'_{\mix} \leq T = O(\diam(G)) =
O(\delta^{-1/2} \log N)$.\qed
\end{proof}

\begin{theorem}[Threshold mixing for $\Z_2^n$ and $K_N$]
The standard transition matrices on the hypercube $\Z_2^n$ and
complete graph $K_N$ satisfy the threshold mixing conjecture.
\end{theorem}
\begin{proof}
We show that Lemma \ref{periodicity} applies in both cases.

For the hypercube:  The eigenvalues of $P$ are $1-\frac{2|k|}{n}$ for
$k \in \Z_2^n$, where $|\cdot|$ is the Hamming weight, so the
eigenvalues of $U_{2\pi n}$ are all $1$; i.e., $U_t$ has period
$2\pi n$ (Moore and Russell \cite{MR}).

For the complete graph:  We may add self-loops (causing a global phase
shift but no loss of generality) and consider the associated
transition matrix $P$.  Its eigenvalues are $1$ (with multiplicity
$1$) and $0$ (with multiplicity $N-1$), so the eigenvalues of
$U_{2\pi}$ are all $1$; i.e., $U_t$ has period $2\pi$.\qed
\end{proof}



\section{Simulability and applications}\label{simulability}

\subsection{Simulation by quantum circuits}

The quantum walk $U_t = e^{-iPt}$ is the continuous-time evolution of
a time-independent Hamiltonian.  This is a valid model of quantum
computation; however, the discrete-time quantum circuit model is often
preferred due to its similarity to the classical digital circuit
model.  We can simulate the continuous-time quantum walk (and
our algorithm) in the circuit model in one of two ways: either by
replacing the walk by its discrete-time counterpart (in which case
simulability is immediate), or by showing that the dynamics of the
walk are well-approximated (i.e., $||U_t - U||_2 \leq \epsilon$) by a
small quantum circuit $U$.\footnote{On the other hand,
the classical continuous-time random walk for time $t$ is trivial to
simulate: simply run the corresponding discrete-time random walk for
$s$ steps, where $s$ is the Poisson random variable with parameter $t$.}
We describe briefly how to do the latter; for more detail, see
Childs \cite{Chi}.

Let $H$ be a Hamiltonian and $\sum_{j=1}^r H_j$ be a decomposition of
$H$ into Hamiltonians $H_j$.  We can simulate $H$ by evolving the
$H_j$ one at a time (in round-robin fashion) by the Lie product formula
\begin{equation}\label{trotter}
e^{-iHt} = (e^{-i H_1 t/j} \cdots e^{-i H_r t/j})^j + O(\max_{k,l}
||[H_k,H_l]||_2 \; r t^2 / j)
\end{equation}
where $[A,B]$ is the commutator $AB-BA$.  Of course, we then have to
worry about simulating each of the $H_j$, but this should be rather
straightforward assuming that we have chosen a good decomposition
$\sum_{j=1}^r H_j$.  For example, if $H=P$ is the standard transition
matrix for the Cayley graph of an Abelian group (product of $d$
cyclic groups), then we consider the natural $2d$-coloring of the
edges $E = \cup_{j=1}^{2d} E_j$ (alternate between two colors along
each of the $d$ directions) and choose each $H_j=P_j$ ($1 \leq j \leq
r = 2d$) such that $P_j(x,y)=P(x,y)$ for all $(x,y) \in E_j$ and
$P_j(x,y)=0$ for all $(x,y) \notin E_j$.  Then each of the $[P_k,P_l]$
is zero and each of the $P_j$ is the disjoint union of locally simulable
interactions, so the overall simulation yields no asymptotic loss in
efficiency.

\subsection{Application to the MCMC method}

In the design of a {\em fully-polynomial randomized approximation scheme}
(FPRAS) for a \#P-complete problem (e.g., the permanent of a matrix
\cite{JSV}), it is often sufficient to construct a
{\em fully-polynomial almost uniform sampler} (FPAUS) over the set of
witnesses (solutions) \cite{Jer}.  Markov chain Monte Carlo (MCMC) algorithms
do this by performing a random walk over the set of witnesses.
The mixing time of this walk shows up in the running time of the
FPRAS; in fact, proving that the walk is {\em rapidly mixing} is
typically the most challenging aspect of designing an FPRAS.  If our
quantum walk algorithm can be shown to offer a speedup over the
classical random walk, it could potentially yield quantum speedups for
a number of FPRAS's.

\section{Conclusions and open problems}\label{conclusions}

We have presented for the first time a natural quantum algorithm based
on the quantum walk $U_t = e^{-iPt}$ that samples from a distribution
$\epsilon$-close to uniform with logarithmic dependence on
$\epsilon^{-1}$ for any symmetric irreducible Markov chain $P$.  We
have outlined a framework for analyzing its running time and
formulated two plausible conjectures which together would imply that
it runs in time $O(\delta^{-1/2} \log N \log \epsilon^{-1})$ when $P$
is the standard transition matrix of a constant-degree graph.  We have
proven each conjecture for a subclass of Cayley graphs.

The most important problems we have left open are the resolution of
the amplification and threshold mixing conjectures.  We note that the
conjectures can be worked on independently, and that partial progress
(as we have obtained for $\Z_p^d$) is indeed possible.  In
particular, we would like to see the threshold mixing conjecture proven
for $\Z_p^d$ or (at least a weak form of) the amplification conjecture
proven for $\Z_2^n$, so that the running time of our algorithm is
determined for $\Z_p^d$ or $\Z_2^n$.  Also, we believe that it should be
possible to prove the amplification conjecture for $\Z_n^d$, $n$
composite, by extending our line of argument.

Another question is whether the decohering walk of Kendon and Tregenna
\cite{KT} (which is similar to our algorithm, only with
Poisson-distributed rather than regular Cesaro-distributed walk
measurements) solves the almost uniform sampling problem with roughly
the same running time as our algorithm; we suspect the answer is yes.
Perhaps it can even be analyzed more easily than our algorithm for
certain classes of walks.

{\bf Acknowledgements.}  I would like to thank Mario Szegedy for many
illuminating discussions on quantum walks and for his help in proving
Lemmas \ref{multiplicities} and \ref{cancellations}.  Thanks also to
Fr\'{e}d\'{e}ric Magniez for his observation in Section \ref{intuition}.

\bibliographystyle{plain}

\begin{thebibliography}{10}

\bibitem{AAKV}
D. Aharonov, A. Ambainis, J. Kempe, U. Vazirani.
\newblock Quantum walks on graphs.
\newblock {\em Proc. STOC}, 2001.  {\tt quant-ph/0012090}

\bibitem{Ald}
D. Aldous.
\newblock Some inequalities for reversible Markov chains.
\newblock {\em Journal of the London Mathematical Society (2)}
25:564-576, 1982.

\bibitem{AF}
D. Aldous, J. Fill.
\newblock Reversible Markov chains and random walks on graphs.
\newblock Book in preparation.

\bibitem{Amb1}
A. Ambainis.
\newblock Quantum walk algorithm for element distinctness.
\newblock {\em Proc. FOCS}, 2004.  {\tt quant-ph/0311001}

\bibitem{Amb2}
A. Ambainis.
\newblock Quantum walks and their algorithmic applications.
\newblock {\em International Journal of Quantum Information}
1:507-518, 2003.  {\tt quant-ph/0403120}

\bibitem{ABNVW}
A. Ambainis, E. Bach, A. Nayak, A. Vishwanath, J. Watrous.
\newblock One-dimensional quantum walks.
\newblock {\em Proc. STOC}, 2001.

\bibitem{AKR}
A. Ambainis, J. Kempe, A. Rivosh.
\newblock Coins make quantum walks faster.
\newblock {\em Proc. SODA}, 2005.  {\tt quant-ph/0402107}

\bibitem{BS}
H. Buhrman, R. Spalek.
\newblock Quantum verification of matrix products.
\newblock {\em Proc. SODA}, 2006.  {\tt quant-ph/0409035}

\bibitem{Chi}
A.M. Childs.
\newblock {\em Quantum information processing in continuous time}.
\newblock Ph.D. Thesis, Massachusetts Institute of Technology, 2004.

\bibitem{CCD}
A.M. Childs, R. Cleve, E. Deotto, E. Farhi, S. Gutmann, D.A. Spielman.
\newblock Exponential algorithmic speedup by quantum walk.
\newblock {\em Proc. STOC}, 2003.  {\tt quant-ph/0209131}

\bibitem{CE}
A.M. Childs, J.M. Eisenberg.
\newblock Quantum algorithms for subset finding.
\newblock {\em Quantum Information and Computation} 5:593, 2005.
{\tt quant-ph/0311038}

\bibitem{CFG}
A.M. Childs, E. Farhi, S. Gutmann.
\newblock An example of the difference between quantum and classical
random walks.
\newblock {\em Quantum Information Processing} 1:35, 2002.  {\tt
quant-ph/0103020}

\bibitem{CG1}
A.M. Childs, J. Goldstone.
\newblock Spatial search by quantum walk.
\newblock {\em Phys. Rev. A} 70:022314, 2004.  {\tt quant-ph/0306054}

\bibitem{CG2}
A.M. Childs, J. Goldstone.
\newblock Spatial search and the Dirac equation.
\newblock {\em Phys. Rev. A} 70:042312, 2004.  {\tt quant-ph/0405120}

\bibitem{CJ}
J.H. Conway, A.J. Jones.
\newblock Trigonometric diophantine equations (On vanishing sums of
roots of unity).
\newblock {\em Acta Arithmetica} 30(3):229-240, 1976.

\bibitem{DS}
P. Diaconis, D. Strook.
\newblock Geometric bounds for eigenvalues of Markov chains.
\newblock {\em Annals of Applied Probability} 1:36-61, 1991.

\bibitem{FG}
E. Farhi, S. Gutmann.
\newblock Quantum computation and decision trees.
\newblock {\em Phys. Rev. A}, 1998.  {\tt quant-ph/9706062}

\bibitem{GW}
H. Gerhardt, J. Watrous.
\newblock Continuous-time quantum walks on the symmetric group.
\newblock {\em Proc. RANDOM}, 2003. {\tt quant-ph/0305182}

\bibitem{Gro}
L.K. Grover.
\newblock A fast quantum mechanical algorithm for database search.
\newblock {\em Proc. STOC}, 1996.  {\tt quant-ph/9605043}

\bibitem{Hal}
S. Hallgren.
\newblock Polynomial-time quantum algorithms for Pell's equation and
the principal ideal problem.
\newblock {\em Proc. STOC}, 2002.

\bibitem{Jer}
M. Jerrum.
\newblock {\em Counting, sampling and integrating: algorithms and complexity.}
\newblock Lectures in Mathematics, ETH Zurich.  Birkhauser, 2003.

\bibitem{JSV}
M. Jerrum, A. Sinclair, E. Vigoda.
\newblock A polynomial-time approximation algorithm for the permanent
of a matrix with non-negative entries.
\newblock {\em J. ACM} 51(4):671-697, 2004.

\bibitem{Kem}
J. Kempe.
\newblock Quantum random walks -- an introductory overview.
\newblock {\em Contemporary Physics} 44(4):307-327, 2003.  {\tt
quant-ph/0303081}.

\bibitem{Kem2}
J. Kempe.
\newblock Discrete quantum walks hit exponentially faster.
\newblock {\em Proc. RANDOM}, 2003.  {\tt quant-ph/0205083}

\bibitem{Ken}
V. Kendon.
\newblock Decoherence in quantum walks - a review.
\newblock {\tt quant-ph/0606016}

\bibitem{KT}
V. Kendon, B. Tregenna.
\newblock Decoherence can be useful in quantum walks.
\newblock {\em Physical Review A} 67:042315, 2003. {\tt
quant-ph/0209005}

\bibitem{Kit}
A.Yu. Kitaev.
\newblock Quantum measurements and the Abelian stabilizer problem.
\newblock {\tt quant-ph/9511026}

\bibitem{MN}
F. Magniez, A. Nayak.
\newblock Quantum complexity of testing group commutativity.
\newblock {\em Proc. ICALP}, 2005.  {\tt quant-ph/0506265}

\bibitem{MSS}
F. Magniez, M. Santha, M. Szegedy.
\newblock An $O(n^{1.3})$ quantum algorithm for the triangle problem.
\newblock {\em Proc. SODA}, 2005.  {\tt quant-ph/0310134}

\bibitem{Mey}
D. Meyer.
\newblock From quantum cellular automata to quantum lattice gases.
\newblock {\em J. Statistical Physics} 85:551-574, 1996.  {\tt
quant-ph/9604003}

\bibitem{MR}
C. Moore, A. Russell.
\newblock Quantum walks on the hypercube.
\newblock {\em Proc. RANDOM}, 2002. {\tt quant-ph/0104137}

\bibitem{NV}
A. Nayak, A. Vishwanath.
\newblock Quantum walk on the line.
\newblock DIMACS TR 2000-43.  {\tt quant-ph/0010117}

\bibitem{SKW}
N. Shenvi, J. Kempe, K.B. Whaley.
\newblock A quantum random walk search algorithm.
\newblock {\em Phys. Rev. A} 67(5):052307, 2003.  {\tt quant-ph/0210064}

\bibitem{Sho}
P. Shor.
\newblock Polynomial-time algorithms for prime factorization and
discrete logarithms on a quantum computer.
\newblock {\em Proc. FOCS}, 1994.  {\tt quant-ph/9508027}

\bibitem{Sim}
D. Simon.
\newblock On the power of quantum computation.
\newblock {\em Proc. FOCS}, 1994.

\bibitem{Sze1}
M. Szegedy.
\newblock Spectra of quantized walks and a $\sqrt{\delta \varepsilon}$-rule.
\newblock {\tt quant-ph/0401053}

\bibitem{Sze2}
M. Szegedy.
\newblock Quantum speed-up of Markov chain based algorithms.
\newblock {\em Proc. FOCS}, 2004.

\bibitem{Wat}
J. Watrous.
\newblock Quantum simulations of classical random walks and undirected
graph connectivity.
\newblock {\em J. Comput. Sys. Sci.} 62(2):376-391, 2001.  {\tt
cs.CC/9812012}

\end{thebibliography}

\end{document}